\documentclass[a4paper,fleqn,usenatbib]{mnras}
\usepackage{footmisc}
\usepackage{graphicx}
\usepackage{amssymb}
\usepackage{amsmath}
\usepackage{wasysym}
\usepackage{color}
\usepackage[normalem]{ulem}
\usepackage{psfrag}
\usepackage{multirow,bigdelim}
\graphicspath{{./}{./plots/}}

\newcommand{\Lya}{\hbox{Ly$\alpha$}}

\newcommand{\halpha}{\hbox{H$\alpha$}}

\newcommand{\hdelta}{\hbox{H$\delta$}}

\newcommand{\fnl}{\hbox{iPTF16fnl}}
\newcommand{\hostname}{\hbox{Mrk 0950}}

\newcommand{\kms}{\hbox{km~s$^{-1}$}}

\newcommand{\mbh}{\hbox{$M_{\rm BH}$}}
\newcommand{\msun}{\hbox{M$_{\odot}$}}
\newcommand{\rsun}{\hbox{R$_{\odot}$}}

\newcommand{\SWIFT}{\textit{Swift}}
\newcommand{\GALEX}{GALEX}

\newcounter{minirefcount}
\title[The UV Evolution of iPTF16fnl]{The Ultraviolet Spectroscopic Evolution of the Low-Luminosity Tidal Disruption Event iPTF16fnl}

\author[J. S. Brown et al.]{J. S. Brown$^{1}$\thanks{E-mail: brown@astronomy.ohio-state.edu},
C. S. Kochanek$^{1,2}$,
T. W.-S. Holoien$^{1,2}$,
K. Z. Stanek$^{1,2}$, \newauthor
K. Auchettl$^{2,3}$,
B. J. Shappee$^{4,5,6}$,
J. L. Prieto$^{7,8}$,
N. Morrell$^{9}$, 
E. Falco$^{10}$, \newauthor
J. Strader$^{11}$,
L. Chomiuk$^{11}$,
R. Post$^{12}$,
S. Villanueva Jr.$^{1}$,
S. Mathur$^{1,2}$, \newauthor
S. Dong$^{13}$,
P. Chen$^{13}$, and
S. Bose$^{13}$\\
$^{1}$ Department of Astronomy, The Ohio State University, 140 West 18th Avenue, Columbus, OH 43210, USA\\
$^{2}$ Center for Cosmology and Astro-Particle Physics, The Ohio State University, 191 West Woodruff Avenue, Columbus, OH 43210, USA\\
$^{3}$ Department of Physics, The Ohio State University, 191 W. Woodruff Avenue, Columbus, OH 43210, USA\\
$^{4}$ Carnegie Observatories, 813 Santa Barbara Street, Pasadena, CA 91101, USA\\
$^{5}$ Hubble Fellow\\
$^{6}$ Carnegie-Princeton Fellow\\
$^{7}$ N\'ucleo de Astronom\'ia de la Facultad de Ingenier\'ia y Ciencias, Universidad Diego Portales, Av. Ej\'ercito 441, Santiago, Chile \\
$^{8}$ Millennium Institute of Astrophysics, Santiago, Chile \\
$^{9}$ Las Campanas Observatory, Carnegie Observatories, Casilla 601, La Serena, Chile \\
$^{10}$ Harvard-Smithsonian Center for Astrophysics, 60 Garden St., Cambridge, MA 02138, USA \\
$^{11}$ Department of Physics and Astronomy, Michigan State University, East Lansing, MI 48824 \\
$^{12}$ Post Observatory, Lexington, MA 02421 \\
$^{13}$ Kavli Institute for Astronomy and Astrophysics, Peking University, Yi He Yuan Road 5, Hai Dian District, Beijing 100871, China
}

\date{Accepted XXX. Received YYY; in original form ZZZ}

\pubyear{2017}

\begin{document}
\label{firstpage}
\pagerange{\pageref{firstpage}--\pageref{lastpage}}
\maketitle

\begin{abstract}
We present the ultraviolet (UV) spectroscopic evolution of a tidal disruption event (TDE) for the first time. After the discovery of the nearby TDE \fnl, we obtained a series of observations with the Space Telescope Imaging Spectrograph (STIS) onboard the Hubble Space Telescope (HST). The dominant emission features closely resemble those seen in the UV spectra of the TDE ASASSN-14li and are also similar to those of N-rich quasars. However, there is significant evolution in the shape and central wavelength of the line profiles over the course of our observations, such that at early times the lines are broad and redshifted, while at later times the lines are significantly narrower and peak near the wavelengths of their corresponding atomic transitions. Like ASASSN-14li, but unlike N-rich quasars, \fnl\ shows neither \ion{Mg}{ii}~$\lambda 2798$\AA\ nor \ion{C}{iii}]~$\lambda 1909$\AA\ emission features. We also present optical photometry and spectroscopy, which suggest that the complex \ion{He}{ii} profiles observed in the optical spectra of many TDEs are in part due to the presence of \ion{N}{iii} and \ion{C}{iii} Wolf-Rayet features, which can potentially serve as probes of the far-UV when space-based observations are not possible. Finally, we use \SWIFT\ XRT and UVOT observations to place strong limits on the X-ray emission and determine the characteristic temperature, radius, and luminosity of the emitting material. We find that \fnl\ is subluminous and evolves more rapidly than other optically discovered TDEs.
\end{abstract}


\section{Introduction}
\label{sec:intro}

When a star passes through the center of a galaxy, it may be disrupted by the central supermassive black hole (SMHB) in a ``tidal disruption event'' (TDE). After the disruption of a main sequence star, approximately half of the stellar debris will remain on bound orbits and asymptotically return to pericenter at a rate proportional to $t^{-5/3}$ \citep{Rees88,Evans89,Phinney89}. While the first TDEs were discovered as X-ray transients \citep[e.g.][]{Grupe99,KomossaBade99,KomossaGreiner99}, the increasing number of wide-field optical transient surveys have led to more recent TDEs being discovered as luminous blue flares from otherwise passive galactic nuclei \citep{Gezari12,Arcavi14,Holoien14}. 

TDEs are a highly inhomogenous class of objects. The optical spectra and X-ray properties in particular are highly diversified \citep[e.g][]{Gezari12,Holoien14,Arcavi14,Vinko15,Holoien16_14li,Holoien16_15oi,Auchettl16}. Furthermore, the evolutionary timescales, peak luminosities, and spectral energy distributions of optical TDEs vary considerably from one TDE to the next \citep[e.g.,][]{Gezari15,Brown16_14ae,Brown16_14li,Holoien16_15oi}. Despite their non-uniform characteristics, all optically discovered TDEs are associated with a hot (tens of thousands of Kelvin) and luminous ($L\gtrsim10^{40}$ ergs s$^{-1}$) flare from an otherwise non-active galactic nuclei. While TDEs are now most freqently discovered in optical transient surveys, energetically speaking, they are predominantly ultraviolet (UV) phenomena. While many TDEs have been monitored with broadband UV filters, UV spectra can reveal much more about the kinematics and ionization structure of the debris. \citet{Cenko16} obtained HST/STIS spectra of what was, at the time, the closest TDE ever discovered \citep[ASASSN-14li;][]{Holoien16_14li}. Their observations revealed the presence of highly ionized, broad C, N, and Si emission, in addition to ionized H and He that had been previously observed in the optical. The lack of \ion{Mg}{ii} emission, a feature universally seen in AGN/quasar spectra, was also particularly interesting.

The transient \fnl\ was discovered on 2016-08-26 and was promptly classified as a TDE on 2016-08-31 \citep{Gezari16}. The position of the transient (J2000 RA/Dec =  00:29:57.04 +32:53:37.5) is consistent with the center of the galaxy \hostname\ at a redshift of $z=0.0163$, corresponding to a luminosity distance of $d = 67.8$~Mpc ($H_0 = 73$~km~s$^{-1}$~Mpc$^{-1}$, $\Omega_M = 0.27$, $\Omega_{\Lambda} = 0.73$). In the discovery paper, \citet{Blagorodnova17} analyzed the photometric evolution and optical spectroscopy and demonstrated that \fnl\ is indeed a rapidly evolving, low luminosity TDE. Our analysis was carried out independently, and is broadly consistent with the results of that paper, though there are some important differences. After \citet{Gezari16} classified the transient as a TDE, we inspected the reported location in ASAS-SN data and found a transient near our sensitivity limit. Since the transient exceeded our triggering criteria, was very nearby, and was apparently brightening, we triggered our HST ToO program and obtained 3 HST/STIS UV spectra of the TDE over the course of the following six weeks. We also obtained \SWIFT\ observations, ground-based photometric monitoring, and optical spectra probing the early and late phases of the event.

In this paper, we present the results of this observing campaign. In Section~\ref{sec:observations} we present UV spectra taken with HST/STIS alongside broadband X-ray, UV, and optical photometry. In Section~\ref{sec:analysis} we use our observational data to model the physical nature of the transient, and also place a stringent upper limit on the X-ray emission from the TDE. In Section~\ref{sec:conclusions} we and summarize our results and discuss the most important implications of our findings.

\section{Observations}
\label{sec:observations}
Once the transient \fnl\ was classified as a TDE, we initiated a follow-up campaign in order to study in detail the nearest TDE yet discovered. In this Section we summarize the observational data used in our analysis.

\subsection{Archival Host Data}

\begin{table}
\centering
\caption{Photometry of \hostname}
\label{tab:arcHostMag}
\begin{tabular}{@{}ccc}
\hline
Filter & Magnitude & Magnitude Uncertainty \\
\hline
$FUV$ & 21.10 & 0.27 \\
$NUV$ & 19.87 & 0.08 \\
$u$ &  17.62 & 0.01 \\
$g$ &  16.08 & 0.01 \\
$r$ &  15.49 & 0.01 \\
$i$ &  15.18 & 0.01 \\
$z$ &  14.92 & 0.01 \\
$J$ &  14.69 & 0.05 \\
$H$ &  14.47 & 0.05 \\
$K_s$ &  14.68 & 0.05 \\

\hline
\end{tabular}

\medskip
\raggedright
\noindent 
These are 5\farcs{0} radius aperture magnitudes from GALEX, SDSS, and 2MASS. 
Magnitudes are in the AB system.
\end{table}

In order to constrain the properties of the host galaxy, we performed a search of archival databases for observations of \hostname. We retrieved archival $ugriz$ images of \hostname\ from the Twelfth Data Release of the Sloan Digital Sky Survey \citep[SDSS;][]{Alam15}. We also retrieved near-IR $JHK_s$ Two-Micron All Sky Survey \citep[2MASS;][]{Strutskie06} images of the host, as well as UV imaging from the Galaxy Evolution Explorer \citep[\GALEX;][]{Martin05}. The AllWISE color of $W1-W2 = 0.01 \pm 0.05$ \citep{Cutri13} indicates that \hostname\ lacks a strong AGN component \citep{Stern12,Assef13}. We examine images from the \textit{ROSAT} All-Sky Survey \citep[RASS;][]{Voges99} and derive an upper limit on the count rate in the 0.3--10~keV band of $<1.9\times10^{-2}$~counts~s$^{-1}$. This corresponds to a limit on the unabsorbed flux of $\sim2\times10^{-13}$~ergs~s$^{-1}$~cm~$^{-2}$ ($L_X \lesssim 1.1\times10^{41}$~ergs s$^{-1}$), which provides further evidence that \hostname\ lacks a strong AGN. We performed aperture photometry on the GALEX, SDSS, and 2MASS images and measured the flux enclosed in a 5\farcs0 aperture. We present the host magnitudes derived from the archival data in Table~\ref{tab:arcHostMag}.

There are no archival Spitzer, Herschel, HST, Chandra, or X-ray Multi-Mirror Mission (XMM-Newton) observations of \hostname, and there are no radio sources listed at this position in the FIRST \citep{Becker95} or NVSS \citep{Condon98} catalogs.

We modeled the host galaxy SED with the Fitting and Assessment of Synthetic Templates \citep[FAST;][]{Kriek09}, using the fluxes derived from the \GALEX, SDSS, and 2MASS archival images. We assumed a \citet{Cardelli89} extinction law with $R_V = 3.1$ and Galactic extinction of $A_V = 0.22$ \citep{Schlafly11}, an exponentially declining star-formation history, a Salpeter IMF, and the \citet{Bruzual03} stellar population models. We obtained a reasonable fit ($\chi^2_{\nu} = 0.98$) which yielded the following parameters: $M_* = (2.34_{-0.05}^{+0.61})\times10^9 \msun$, age$ = 1.29_{-0.03}^{+0.33}$ Gyr, and a 1-$\sigma$ upper limit on the star formation rate of SFR$<0.007$~\msun~yr$^{-1}$. 

There are no black hole mass measurements for this galaxy in the literature. In order to obtain an estimate of the black hole mass, we adopt an upper limit on the bulge mass using the stellar mass derived from the photometry in the central 5$\arcsec$ of the galaxy. This a reasonable approximation given the properties of the light profile measured from the SDSS imaging (i.e. half light radius $\sim3\farcs5$). We then adopt the $M_{\rm bulge}$--$M_{\rm BH}$ scaling relation from \citet{McConnell13} and estimate the mass of the black hole to be $M_{\rm BH} \lesssim 5.5\times10^6$\msun. Thus the black hole is not so massive that a main sequence star would simply be absorbed by the SMBH before being tidally disrupted. Given the lack of archival spectra, we will address the spectroscopic nature of the host in Section~\ref{sec:analysis}.

\subsection{ASAS-SN Detection}
After the discovery of the transient was announced \citep{Gezari16}, we examined the location of the object in ASAS-SN data. The last ASAS-SN epoch before discovery was observed on UT 2016-08-26.51 under moderate conditions by the quadruple 14-cm ``Brutus'' telescope in Haleakala, Hawaii. This ASAS-SN field was processed by the standard ASAS-SN pipeline \citep{Shappee14} using the \textit{isis} image subtraction package \citep{Alard98,Alard00}, except we took extra care to exclude images with flux from \fnl\ from the construction of the reference image. We performed aperture photometry at the location of \fnl\ on the subtracted images using the \textsc{iraf} \textit{apphot} package and calibrated the results using the AAVSO Photometric All-Sky Survey \citep[APASS;][]{Henden16}. We have detections of the transient on UT 2016-08-26.51 ($V=17.96\pm0.24$), and on UT 2016-08-30.52 ($V=17.49\pm0.18$). We do not detect the object in images taken before UT 2016-08-26.51, but our limits preceeding the discovery of the transient are relatively shallow ($V\gtrsim17$), so this is not to say the transient is not present at earlier epochs.

\subsection{HST/STIS ToO Observations}

We observed iPTF16fnl using STIS and the FUV/NUV MAMA detectors. We used the $52\farcs0 \times 0\farcs2$ slit and the G140L (1425\AA, FUV-MAMA) and G230L (2736\AA, NUV-MAMA) gratings.  The visits were obtained over 1, 2 and 4 orbits, respectively, as the target faded.  The integration times were 792 seconds (FUV and NUV) in four equal exposures (each setting) for the first visit, 1854 (FUV) and 2466 (NUV) seconds in six exposures for the second visit, and 4785  (FUV) and 5500 (NUV) seconds in ten roughly equal exposures for the third visit.  A cosmic ray CR-SPLIT image pair was obtained at each dither position and the telescope was dithered along the slit by 16 pixels. Since the trace of \fnl\ was clearly present in the two-dimensional frames and spatially unresolved, we use the one-dimentional spectra output by the HST pipeline. For each epoch, we perform inverse-variance-weighted combinations of the one-dimensional spectra and merge the FUV and NUV channels. We correct for Galactic reddening assuming $A_V = 0.22$ and $R_V = 3.1$ \citep{Odonnell94,Schlafly11}. The merged spectra are shown in Figure ~\ref{fig:nuvSpec}.

\subsection{\SWIFT\ Observations}

We also obtained \SWIFT\ observations of \fnl. The UVOT \citep{Roming05} observations were obtained in six filters: $V$ (5468 \AA), $B$ (4392 \AA), $U$ (3465 \AA), $UVW1$ (2600 \AA), $UVM2$ (2246 \AA), and $UVW2$ (1928 \AA). We used the UVOT software task \textsc{uvotsource} to extract the source counts from a 5\farcs0 radius region and a sky region with a radius of $\sim$~40\arcsec. The UVOT count rates were converted into magnitudes and fluxes based on the most recent UVOT calibration \citep{Poole08, Breeveld10}. The observed UVOT magnitudes are shown in Figure~\ref{fig:swiftLC} and are tabulated in Table~\ref{tab:phot}.

We simultaneously obtained \SWIFT\ X-ray Telescope \citep[XRT;][]{Burrows05} observations of the source. The XRT was run in photon-counting (PC) mode \citep{Hill04} which is the standard imaging mode of the XRT. We reduced all observations following the \SWIFT\ XRT data reduction guide\footnote{\url{http://swift.gsfc.nasa.gov/analysis/xrt\_swguide\_v1\_2.pdf}} and reprocessed the level one XRT data using the \SWIFT\ \textit{xrtpipeline} version 0.13.2 script, producing cleaned event files and exposure maps for each observation.

To extract the number of background subtracted source counts in the 0.3--10.0~keV energy band from each individual observation, we used a source region centered on the position of \fnl\ with a radius of $50\arcsec$ and a source free background region centered at $(\alpha,\delta)=(00^{h}29^{m}25.5^{s}, +32^{\circ}51'15.6'')$ with a radius of 200$\arcsec$. We correct for Galactic absorption assuming a column density of $5.6\times10^{20}$~cm$^{-2}$ \citep{Kalberla05}.

\subsection{Ground-Based Monitoring}

In addition to space-based observations, we also obtained ground based photometric and spectroscopic monitoring of \fnl\ for approximately 4 months following discovery. The photometric monitoring consisted of $BgVri$ imaging, which we obtained from several telescopes including 1-m telescopes on the Las Cumbres Observatory telescope network \citet{Brown13}, the 20-in DEMONEXT telescope \citep{Villanueva16}, and the 24-in telescope at Post Observatory. We measured aperture photometry with the same 5\farcs0 aperture radius used for both the \SWIFT\ and archival host photometry. Photometric zero-points were determined using nearby stars with SDSS and/or APASS magnitudes.

We also obtained several early ($t \lesssim 14$ days) optical spectra of \fnl, as well as late-time spectra taken after the transient had faded considerably. The early time spectra were obtained with the FAST Spectrograph \citep[FAST;][]{Fabricant98} on the Fred L. Whipple Observatory Tillinghast 1.5-m telescope, the Wide Field Reimaging CCD Camera (WFCCD) mounted on the Las Campanas Observatory du Pont 2.5-m telescope, and the Goodman Spectrograph on the Southern Astrophysical Research (SOAR) 4.1-m telescope. The late-time spectra were obtained with the Multi-Object Double Spectrographs \citep[MODS;][]{Pogge10} mounted on the dual 8.4-m Large Binocular Telescope (LBT) on Mt. Graham. The MODS data were reduced with a combination of the \textsc{modsccdred}\footnote{\url{http://www.astronomy.ohio-state.edu/MODS/Software/modsCCDRed/}} \textsc{python} package and the \textsc{modsidl} pipeline\footnote{\url{http://www.astronomy.ohio-state.edu/MODS/Software/modsIDL/}}. The other spectroscopic data were reduced using standard techniques in \textsc{iraf}. While there are slight difference in the wavelength coverage and resolution of the spectrographs, the spectra cover the majority of the optical bandpass ($\sim3000$\AA\ to $\gtrsim7000$\AA) and have a resolution of $\sim5$\AA. Finally, in order to facilitate comparison of spectra across multiple observing epochs and instruments, we flux-calibrated each spectra with contemporaneous $r$-band imaging.

\section{Analysis}
\label{sec:analysis}

In this section we present the evolution of \fnl\ from $\sim1200$\AA\ up to $\sim8000$\AA. This is the first time the temporal evolution of a TDE has been spectroscopically observed in the the UV, and provides key constraints for theories attempting to explain the observed properties of TDEs. 

\subsection{Spectroscopic Analysis}
\label{sec:specAnalysis}

\begin{figure*}
\centering{\includegraphics[scale=1.,width=\textwidth,trim=0.pt 0.pt 0.pt 0.pt,clip]{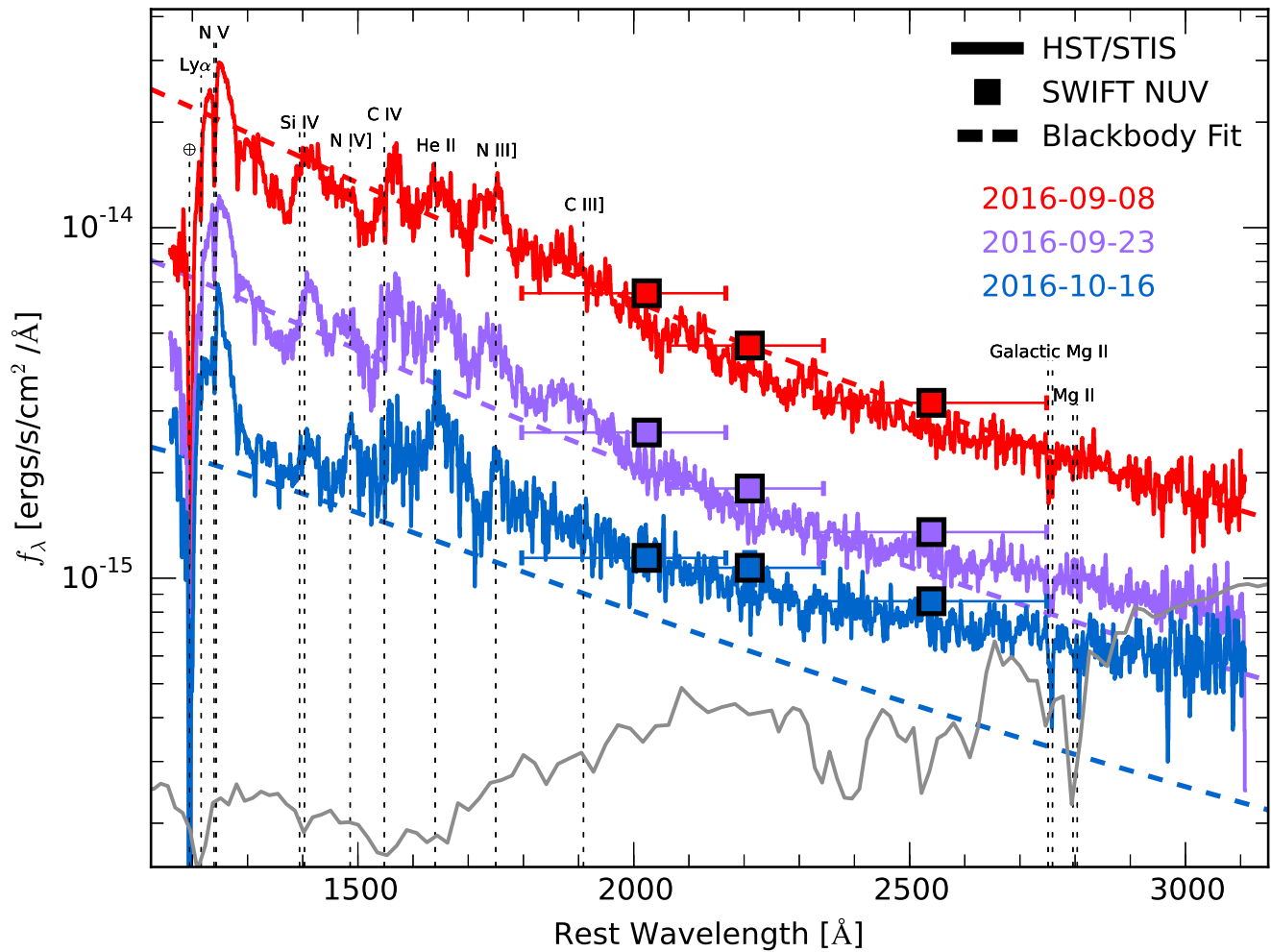}}
\caption{The UV evolution of \fnl\ as revealed by HST/STIS spectra and \SWIFT\ photometry. The spectra have been smoothed with a 5 pixel boxcar and scaled by a constant factor to best match the \SWIFT\ photometry for ease of comparison. The dashed lines show our blackbody fits to the host subtracted \SWIFT\ fluxes. Prominent atomic transitions are marked with vertical dotted lines. The thin gray line shows our estimate of the UV spectrum of the host based on the SED model.}
\label{fig:nuvSpec}
\end{figure*}

\begin{figure*}
\centering{\includegraphics[scale=1.,width=\textwidth,trim=0.pt 0.pt 0.pt 0.pt,clip]{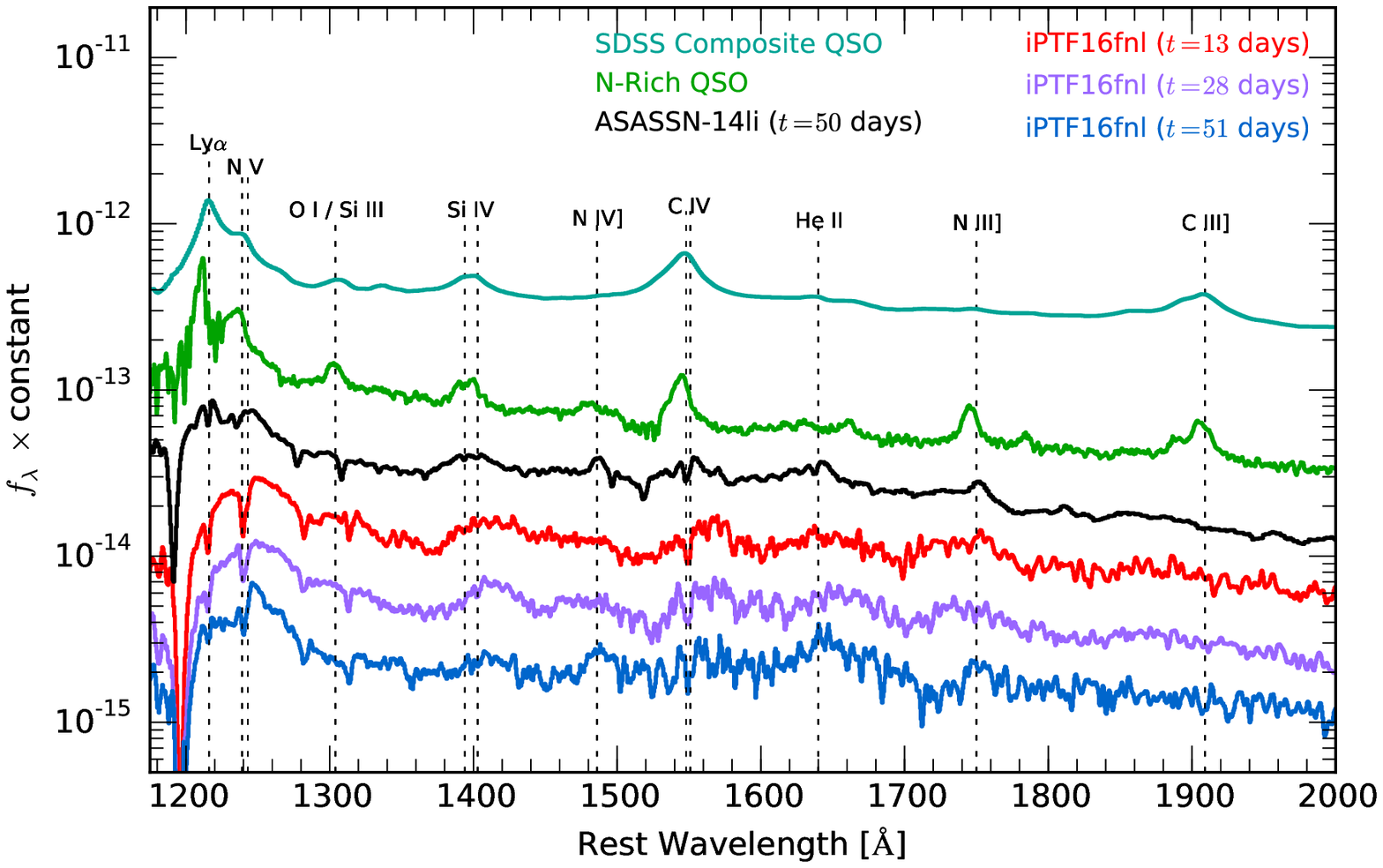}}
\caption{Comparison of the HST/STIS spectral evolution of \fnl\ (colored lines) with the HST/STIS spectrum of ASASSN-14li from \citet{Cenko16} (black), a nitrogen rich quasar \citep{Batra14}, and the composite quasar spectrum of \citet{vandenBerk01}. Spectra have been offset by a constant factor for clarity. Prominent features associated with atomic transitions are labeled. Each spectrum is shown in the rest frame of the respective object. The Galactic absorption features appearing in both \fnl\ and ASASSN-14li spectra appear at bluer wavelengths in the spectrum of ASASSN-14li due to the different redshifts of the objects.}
\label{fig:nuvSpecLI}
\end{figure*}

\begin{figure*}
\centering{\includegraphics[scale=1.,width=\textwidth,trim=0.pt 0.pt 0.pt 0.pt,clip]{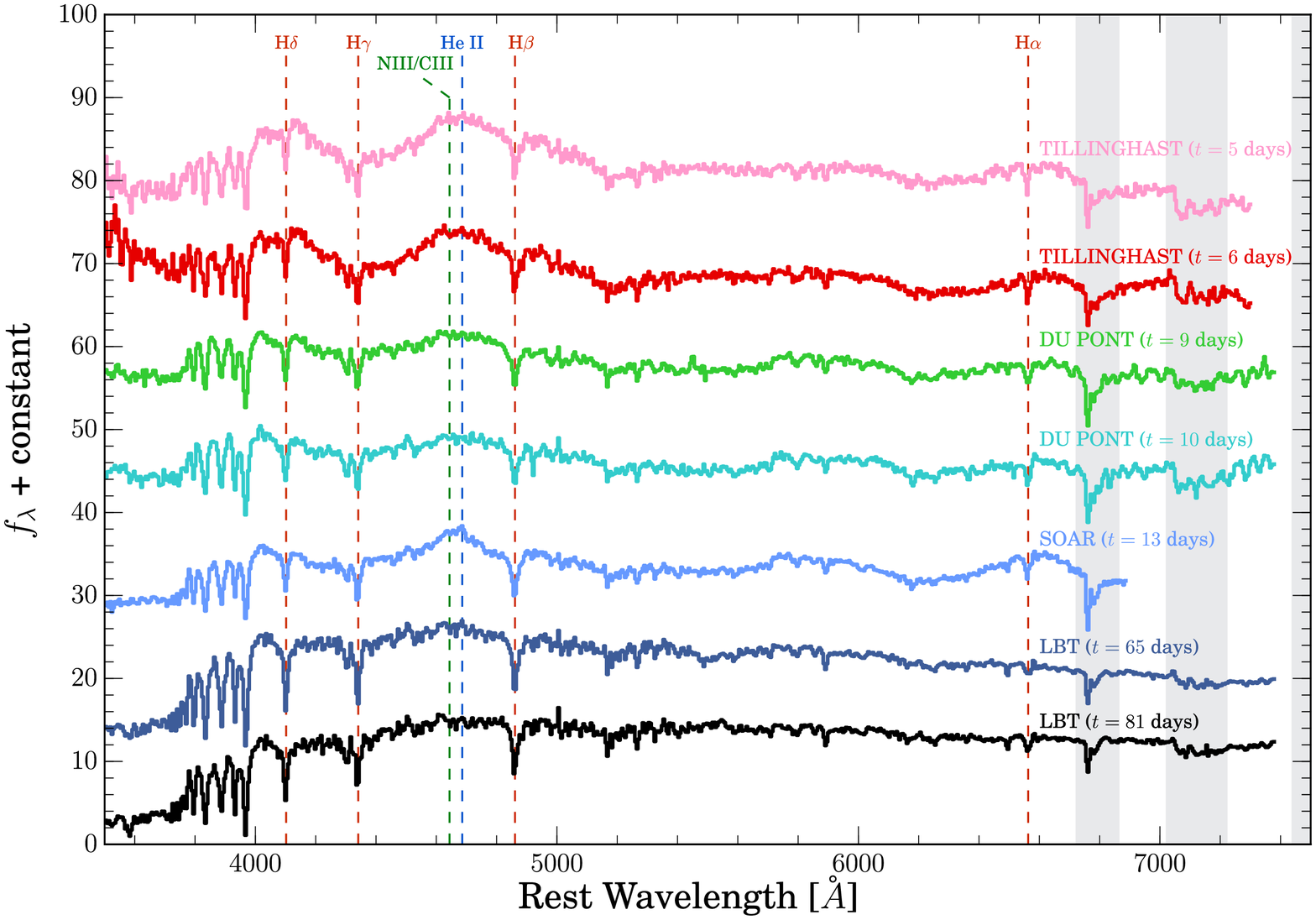}}
\caption{Optical spectral evolution of \fnl. Color denotes days since discovery. Prominent atomic transitions are labeled, and the shaded regions show the location of telluric features where systematic errors may be significant. The strength of the blue continuum and emission line features decrease with time. The latest spectrum taken with MODS/LBT closely resembles that of a post-starburst galaxy; there is little evidence for residual TDE emission.}
\label{fig:opticalSpec}
\end{figure*}

The spectroscopic evolution of \fnl\ in the UV is shown in Figure~\ref{fig:nuvSpec}. Each spectrum is shown in a color corresponding to the observation date. The squares show the contemporaneous \SWIFT\ observations, which, for purposes of comparison, we have used to set the overall normalization of the HST/STIS spectra. The gray line shows the best fit SED based on the archival host data, while the dashed curves show our blackbody fits to the host-subtracted UVOT fluxes (see Section~\ref{sec:photAnalysis}). Prominent atomic transitions are labeled, and the portion of the spectrum affected by geocoronal airglow is marked with the telluric symbol. 

We emphasize that the HST/STIS and \SWIFT\ observations sample different regions of the galaxy and thus have different levels of host contamination. The spectrum is dominated by the blue emission of the TDE at the very center of the galaxy, while the modeling of the host is based on the central 5\arcsec\ of the galaxy. Thus a direct comparison between the two is not possible since the contribution of the host in the HST spectra is not precisely known. However, a relative comparison between the observed \SWIFT\ fluxes, the archival host SED, and our blackbody fits is fair and demostrates the relative contribution of the host and TDE to the observed broadband flux at each epoch.

Our first spectrum of \fnl\ bears a strong resemblance to that of ASASSN-14li from \citet{Cenko16}. Both objects are well approximated by a combination of blackbody continuum emission with a temperature of a few $10^4$~K, coupled with broad emission and narrow absorption associated with highly ionized atomic transitions. Our blackbody fits agree well with the shape of the UV spectra, and the best agreement is seen at earlier times when the flux from the TDE dominates the observed UV emission and the contribution from the underlying host is negligible. At later times, as the transient fades, the observed (host+TDE) spectrum begins to diverge from our blackbody fits because the host contribution (which we subtract before modeling the excess emission) becomes more significant. 

The line profiles are complex; absorption and emission features of multiple lines are blended together and also frequently coincide with Galactic absorption features. In the earlier spectra, the emission features are systematically redshifted relative to the line center. For instance, the peak of the \ion{N}{v} feature is redshifted by $\sim2000-2500$~\kms, as is that of \ion{C}{iv}. Some lower ionization features (e.g. \ion{N}{iii}]) show no apparent redshift at all. The \ion{He}{ii} profile is particularly interesting. While the emission peaks near the transition wavelength at 1640\AA, there is a significant component of the line profile that extends to redder wavelengths. This suggests that the distribution of the highly ionized metals may differ from that of the gas producing the bulk of the lower ionization features. However, given the complex nature of the line blending, it is difficult to differentiate between this scenario and one in which the high ionization lines are systematically broader and/or happen to coincide with more absorption features at shorter wavelengths. In fact, the \ion{Si}{iv} and \ion{C}{iv} show clear blueshifted absorption troughs ($\Delta v\sim$~FWHM~$\sim 6000$~\kms) that may signify an outflow, and also likely contribute to the apparent redshift of the emission line profiles. Additionally, there is a non-negligible amount of Galactic extinction along the line of sight ($A_V = 0.22$), and  nearly all identified line profiles in Figures~\ref{fig:nuvSpec} and~\ref{fig:nuvSpecLI} have relatively strong Galactic absorption features located blueward of the transition wavelength in the rest frame of the TDE, as was also observed by \citet{Cenko16} for ASASSN-14li.

The shapes of the line profiles vary with time. This includes the apparent width of the lines, as well as their peak wavelengths. The systematic redshift observed in \ion{N}{v}, \ion{Si}{iv}, and \ion{C}{iv} at early times is less dramatic in the last HST/STIS spectrum, and the broad absorption features are also less apparent at late times. There is an apparent decrease in flux in the region between \Lya\ and \ion{N}{v}~$\lambda 1240$\AA. While this region is affected by geocoronal afterglow, the latest spectrum shows a substantial drop in flux in this region relative to the previous two. We also note the sudden appearance of the \ion{N}{iv}]~$\lambda 1486$\AA\ line, which was relatively weak in the initial HST/STIS spectrum but quite prominent in the last spectrum. Similarly, the \ion{He}{ii} line becomes much stronger at later times relative to the apparent continuum, and by $\sim50$ days the profile is strongly peaked at the transition wavelength. The evolution of the spectra can be explained in large part by the narrowing of the emission lines as the flare fades. This is opposite the behavior seen in actively accreting SMBHs \citep{McGill08,Denney09}, but appears to be typical of TDEs \citep[e.g.,][]{Holoien14, Holoien16_14li, Holoien16_15oi, Brown16_14ae, Brown16_14li}. The increasing equivalent width of the \ion{He}{ii} feature indicates that it is not fading as fast as the other lines. This may also suggest that the bulk of the \ion{He}{ii} emission is produced by gas that evolves on a longer timescale than the gas producing the highly ionized metal lines. The wavelengths of the narrow absorption features remain constant, while the velocity offsets and widths of the broad absorption features appear to decrease with time. This likely indicates that material responsible for the narrow absorption features is located at larger distances and varies on significantly longer timescales than the material responsible for the broad features. 

Figure~\ref{fig:nuvSpecLI} shows the blue portion of the \fnl\ HST/STIS spectrum compared to that of ASASSN-14li from \citet{Cenko16}. The spectra have been scaled by a constant factor for ease of comparison. The phase of the ASASSN-14li spectrum most closely matches that of our last HST spectrum, though the phases of the two are not necessarily comparable given the uncertainty on the rise times. The spectra are shown in the rest frame of the respective TDEs. Due to the different redshifts of the two objects, Galactic absorption features in the ASASSN-14li spectrum are located at systematically shorter wavelengths than their counterparts in the \fnl\ spectrum. This includes the geocoronal afterglow which strongly affects the \Lya\ portion of the \fnl\ spectrum, but has a subtler effect on the ASASSN-14li spectrum. Nonetheless, both spectra show the same set of prominent emission features. The line profiles of ASASSN-14li most closely resemble those in the latest \fnl\ spectrum, which is not necessarily surprising given the timing of the observations. Specifically, both the ASASSN-14li spectrum and our latest spectrum of \fnl\ show little evidence of the strongly redshifted emission profiles that are so apparent in the early time \fnl\ spectrum.

Interestingly, neither ASASSN-14li nor \fnl\ show the \ion{Mg}{ii} emission typically associated with accreting SMBHs. The \ion{Mg}{ii} emission comes from the relatively large partially ionized regions in AGN \citep[e.g.][]{Peterson93,Richards02}. The absence of \ion{Mg}{ii} emission in TDEs could be a symptom of the shape of the radiation field, or it could indicate that the nebula is matter bounded and the Mg is predominantly in higher ionization states. The ionization energy of \ion{Mg}{ii} is low enough ($E_{\rm ion} = 15.03$~eV) for this to be plausible. Like \ion{Mg}{ii}, \ion{He}{i} emission is typically quite strong in the spectrum of quasars, but is relatively weak in the spectra of TDEs. In the SDSS quasar composite spectrum from \citet{vandenBerk01}, the \ion{He}{i}~$\lambda 5877$\AA\ flux is a factor of a few larger than the \ion{He}{ii}~$\lambda 4686$\AA\ flux. However, ASASSN-14li and \fnl\ show evidence for only weak \ion{He}{i} emission, which has a slightly higher ionization energy than \ion{Mg}{ii} ($E_{\rm ion} = 24.59$~eV). We also note that there is no \ion{C}{iii}] emission in any of the TDE spectra. This line is generally used to constrain the density of the broad line region in AGN to $n\lesssim10^9$~cm$^{-3}$ \citep{Ferland92,Peterson97}, since at higher densities the line is suppressed by collisional deexcitation. The absence of this feature from the TDE spectra may be due to the high density of the emitting material, which would have direct implications for the modeling of TDEs.

Despite these differences between the QSO and TDE spectra, the N-rich QSO spectrum from \citet{Batra14} bears some striking similarities to the UV spectra of TDEs. While this has been previously noted for the UV spectrum of ASASSN-14li \citep{Kochanek16_stellar,Cenko16}, the strong N features are also present in the spectra of \fnl. The prevalence of the N lines in the spectra of TDEs further suggests a link between the population of N-rich QSOs and TDEs \citep{Kochanek16_stellar}.

The optical spectra of \fnl\ taken at $t$ = 9, 10, 13, 65, and 81 days are shown in Figure~\ref{fig:opticalSpec}. The spectra are qualitatively similar to observations of other TDEs \citep[e.g.][]{Arcavi14,Holoien14,Brown16_14ae}, which show a strong blue continuum and broad emission features associated with H and He recombination. \citet{Blagorodnova17} also present optical spectra of \fnl\ that show these same features typically associated with TDEs. The emission line profiles in \fnl\ are relatively broad, the host galaxy is bright, and the TDE itself is faint, causing the optical lines to appear relatively weak. In the spectrum taken at $t=13$ days the \ion{He}{ii}~$\lambda 4686$\AA\ feature is quite prominent. This change may in part be due to a narrowing of the line profile, but it is most easily explained by varying levels of host contamination between the observations and across multiple instruments.

While it is challenging to cleanly separate the TDE emission features in our optical spectra, the \ion{He}{ii} feature is broad and has a substantial amount of flux blueward of 4686\AA. This is most evident in the spectrum taken $\sim13$ days after discovery and is also quite evident in several spectra from \citet{Blagorodnova17}. The origin of these two components not obvious. The most straightforward interpretation is that both features arise from \ion{He}{ii}~$\lambda 4686$ with one component blueshifted by $\sim3000$~\kms\ relative to a component centered at the wavelength of the \ion{He}{ii} transition. The structure of the stellar debris in TDEs is not well understood, and the kinematics inferred from the line profiles may be useful for constraining the distribution of of the emitting material. \citet{Arcavi14} was able to fit the asymetric double peaked \halpha\ profile of the TDE PTF09djl using a Keplerian disk model from \citet{Strateva03}. However, such a model cannot explain the \ion{He}{ii} profile in \fnl\ because the emission lines in the UV (including \ion{He}{ii}) do not show the kinematic structure associated with such a disk.

Other TDEs have shown apparent double peaked \ion{He}{ii} profiles. The most striking example of this is seen in the spectra of ASASSN-14li \citep{Holoien16_14li}. The spectra have two strongly peaked line profiles near \ion{He}{ii}, at approximately $4686$\AA\ and $4640$\AA. In fact, in the earliest spectra of ASASSN-14li, the feature at $\sim4640$\AA\ is significantly stronger than the the feature at $\sim4686$\AA. After $\sim60$ days, the anomalous blue feature fades rapidly, and the component at $4686$\AA remains the stronger of the two for several hundred days \citep{Brown16_14li}. Similarly, the H-poor TDE PS1-10jh \citep{Gezari12,Gezari15} also had a \ion{He}{ii}~$\lambda 4686$\AA\ profile with a significant blue wing, but in that case the peak of the second component was much bluer ($\sim4470$\AA) than that observed in the TDEs discussed here. Evidence for this second emission feature is seen in other TDEs as well (e.g. ASASSN-14ae, PTF09ge), but its presence is more ambiguous than in ASASSN-14li.

Given the new information afforded by the HST/STIS UV spectra, it seems likely that the feature at $\sim4640$\AA\ is in fact due to highly ionized metal lines associated with \ion{C}{iii}, \ion{N}{iii}, and possibly \ion{N}{v} as well. \citet{Gezari15} suggested that the blue wing in the \ion{He}{ii} profile of PS1-10jh could be due to blue shifted \ion{C}{iii}/\ion{N}{iii} Wolf-Rayet blends \citep{Niemela85,Massey98,Leonard00}. These lines (including \ion{N}{v}) are also prominent in the spectra of the flash ionized material surrounding core collapse supernovae \citep{Shivvers15, GalYam14, Khazov16, Yaron17}. The complexity of the \ion{He}{ii} profiles of most optically discovered TDEs suggests that these \ion{C}{iii} and \ion{N}{iii} features are actually quite common, and could exacerbate the apparent He/H ratios observed in many TDEs. These features may serve as useful probes of the UV emission when space-based UV observations are not possible, but rapid spectroscopic follow-up of TDEs in both the UV and optical is needed to precisely characterize how they relate to the emission features in the UV.

It is now clear that TDEs frequently have emission features associated with highly ionized metals. In AGN, the highly ionized lines (e.g. \ion{C}{iv}) typically have blueshifted line profiles \citep[e.g.][]{vandenBerk01,Richards02}, and this is clearly seen in the composite quasar spectrum in Figure~\ref{fig:nuvSpecLI}. The apparent blue shift is thought to arise from a relative absence of red flux due to the geometry of AGN and resulting opacity effects. This systematic blueshift is not observed in \fnl; the highly ionized metal lines are redshifted relative to the systemic velocity. In fact, the emission line profiles in TDEs show a variety of velocity offsets that evolve on a relatively short timescale \citep[e.g.][]{Brown16_14ae,Brown16_14li}. Additionally, ASASSN-15oi appeared to show varying Doppler shifts between different ionic species \citep{Holoien16_15oi}. Thus the varying Doppler shifts of the lines are likely to be partially associated with the geometry of the event and the resulting kinematics of the emitting material, rather than being driven only by opacity effects. Such rapid kinematic evolution is not surprising. For example, material at $r=10^{15}r_{15}$~cm from a SMBH with mass $M_{BH} = 10^6M_{\rm BH6}$~\msun\ has a characteristic velocity of $v=(GM/r)^{1/2} \sim 3000 M_{\rm BH6}^{1/2}r_{15}^{-1/2}$~\kms, and the timescale for gravitational accelerations to significantly change the velocity is of order $t \sim 33 r_{15}^{3/2} M_{\rm BH6}^{-1/2}$ days.

When the late-time MODS spectra were taken the TDE had largly vanished in the optical, and the resulting spectra show no clear signs of TDE emission. The spectra closely resemble that of a post-starburst galaxy \citep[e.g.,][]{Dressler83, Zabludoff96}. The \hdelta\ absorption is quite strong (EW$_{\rm H_{\delta}} \sim-5$\AA), and there are no signatures of significant ongoing star formation. This is fully consistent with the SED modeling, and places this galaxy firmly in the E+A classification. Thus, \fnl\ serves as an additional piece of evidence that TDEs are preferentially found in post-starburst galaxies \citep{French16a,French16b}.

The spatial resolution of our optical spectra is limited to $\sim1\farcs2$ due to a combination modest observing conditions and slit widths. However, the configuration of the most recent LBT/MODS observation on 2016-11-18 allowed us to align the slit with the major axis of the host galaxy and perform multiple extractions along the slit out to $\pm6\arcsec$ (corresponding to a physical scale of $\sim2$~kpc). In the spectra extracted near the galaxy center (Figure~\ref{fig:opticalSpec}), there is relatively weak \ion{O}{iii}~$\lambda 5007$\AA\ emission line. The strength of this nebular feature relative to the stellar continuum increases with distance from the center of the galaxy. Beyond $\sim2$~kpc, the S/N of both the stellar continuum and nebular emission decrease rapidly. While the stellar absorption features show clear velocity structure due to the rotation of the disk, there is very little evidence of any kinematic signature in the emission profiles associated with the nebular gas. The LBT/MODS spectrum from 2016-11-02, despite being misaligned with the major axis of the galaxy by $\sim 45$ degrees, also shows spatially extended nebular emission. While the S/N is relatively low, it appears that the nebular emission probed by this slit orientation does in fact show kinematic structure. It is not unprecedented for the kinematic signatures of the gas to differ from that of a galaxy's stellar population \citep[e.g.][]{Sarzi06,Davis11,Jin16}. Similarly, \citet{Prieto16} showed the host of ASASSN-14li to have spatially extended, filamentary nebular emission well outside the optical extent of the galaxy. In the case of ASASSN-14li, the origin and structure of the nebular gas is most likely related to a merger event \citep{Prieto16,RomeroCanizales16}, which may explain why TDEs are preferentially found in post-starburst galaxies \citep{French16a,French16b}. Future integral field observations of TDE hosts are necessary to characterize the prevalence of disturbed nebular gas around these unusual galaxies.

\subsection{Photometric Analysis}
\label{sec:photAnalysis}

\begin{figure*}
\centering{\includegraphics[scale=1.,width=\textwidth,trim=0.pt 0.pt 0.pt 0.pt,clip]{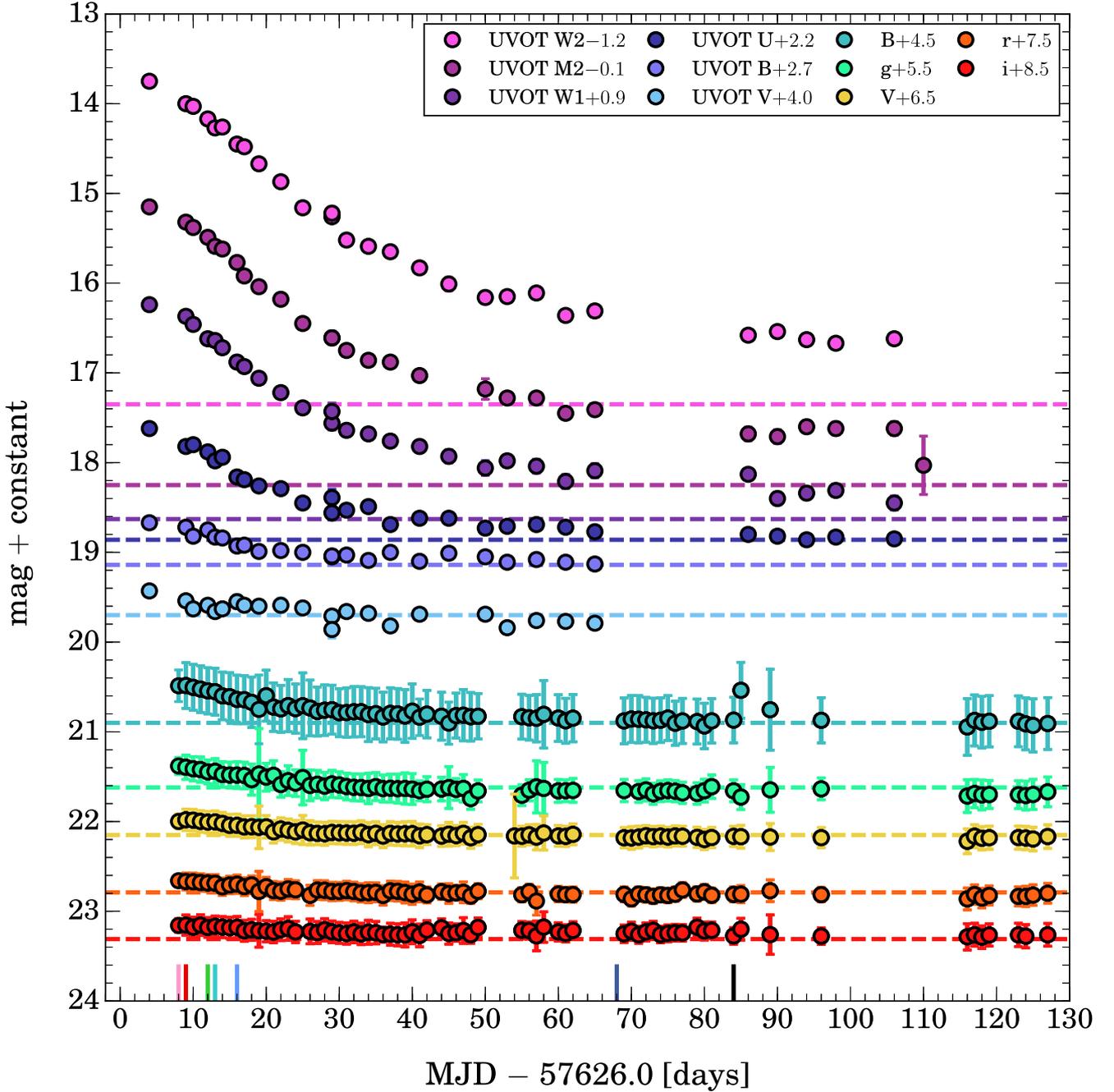}}
\caption{Photometric evolution of \fnl\ from discovery to $\sim 130$ days after discovery. Circles show the observed non-host-subtracted magnitudes. All UV and optical magnitudes are shown in the Vega system. Horizontal dashed lines show the host magnitudes synthesized from the best fit SED. Like other TDEs, \fnl\ remains bright in the UV for significantly longer than in the optical bands, with the bluest bands showing the largest residual flux.}
\label{fig:swiftLC}
\end{figure*}

\begin{figure}
\centering{\includegraphics[scale=1.,width=0.5\textwidth,trim=0.pt 0.pt 0.pt 0.pt,clip]{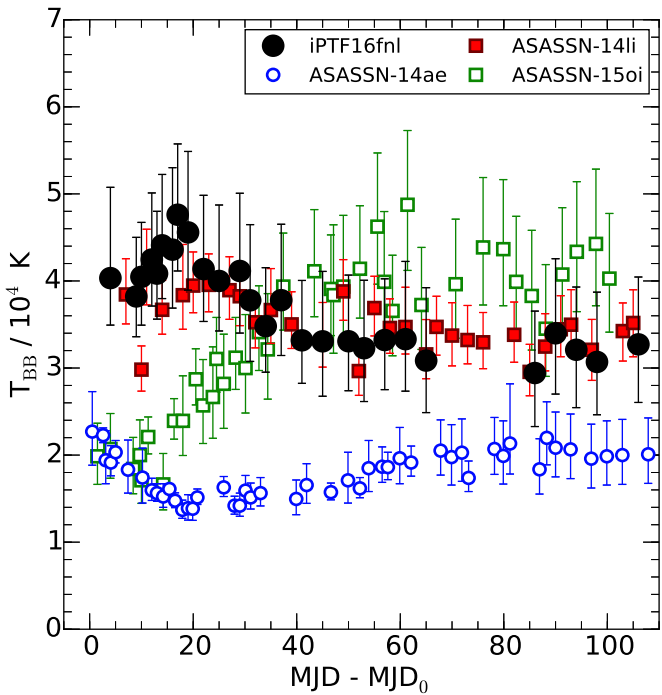}}
\caption{Temperature evolution of \fnl. Black circles show the temperature of the UV/optical continuum of \fnl\ inferred from our blackbody fits to the host subtracted UVOT fluxes. The open blue circles, filled red squares, and open green squares show the temperature evolution of ASASSN-14ae, ASASSN-14li, and ASASSN-15oi, respectively.}
\label{fig:bbTemp}
\end{figure}

\begin{figure}
\centering{\includegraphics[scale=1.,width=0.5\textwidth,trim=0.pt 0.pt 0.pt 0.pt,clip]{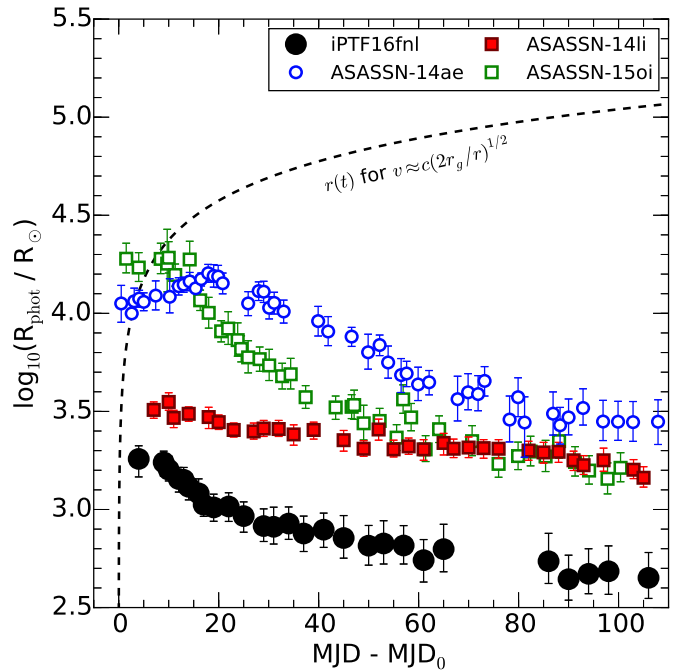}}
\caption{Radius evolution of \fnl. Black circles show the bolometric luminosity of \fnl\ inferred from our blackbody fits to the host subtracted UVOT fluxes; the open blue circles, filled red squares, and open green squares show the radius evolution of ASASSN-14ae, ASASSN-14li, and ASASSN-15oi, respectively. The dashed line shows the evolution of a parabolic orbit with a closest approach equal to the tidal radius for a $10^7 \msun$ SMBH. The optical photosphere of \fnl\ is significantly smaller than those of the three ASAS-SN TDEs.}
\label{fig:bbRad}
\end{figure}

\begin{figure}
\centering{\includegraphics[scale=1.,width=0.5\textwidth,trim=0.pt 0.pt 0.pt 0.pt,clip]{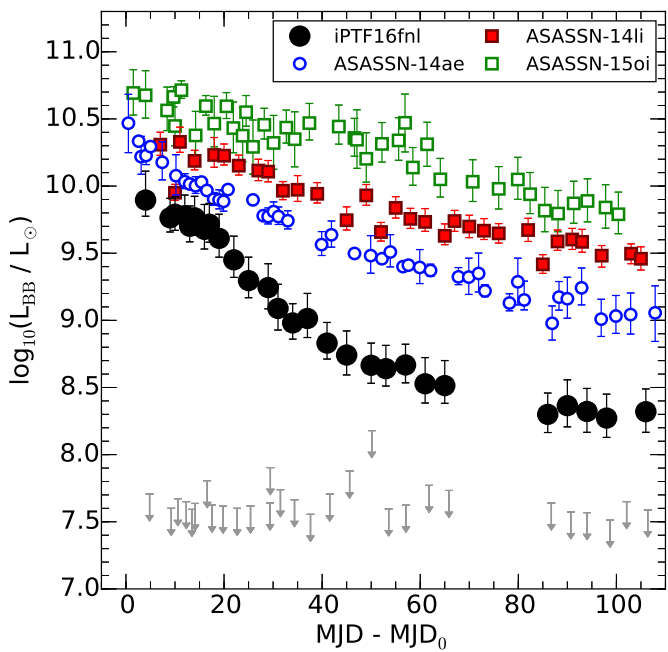}}
\caption{Luminosity evolution of \fnl. Black circles show the bolometric luminosity of \fnl\ inferred from our blackbody fits to the host subtracted UVOT fluxes; the open blue circles, filled red squares, and open green squares show the temperature evolution of ASASSN-14ae, ASASSN-14li, and ASASSN-15oi, respectively. The bolometric luminosity of \fnl\ is significantly lower than that of the ASAS-SN TDEs.}
\label{fig:bbLum}
\end{figure}

\begin{figure}
\centering{\includegraphics[scale=1.,width=0.5\textwidth,trim=0.pt 0.pt 0.pt 0.pt,clip]{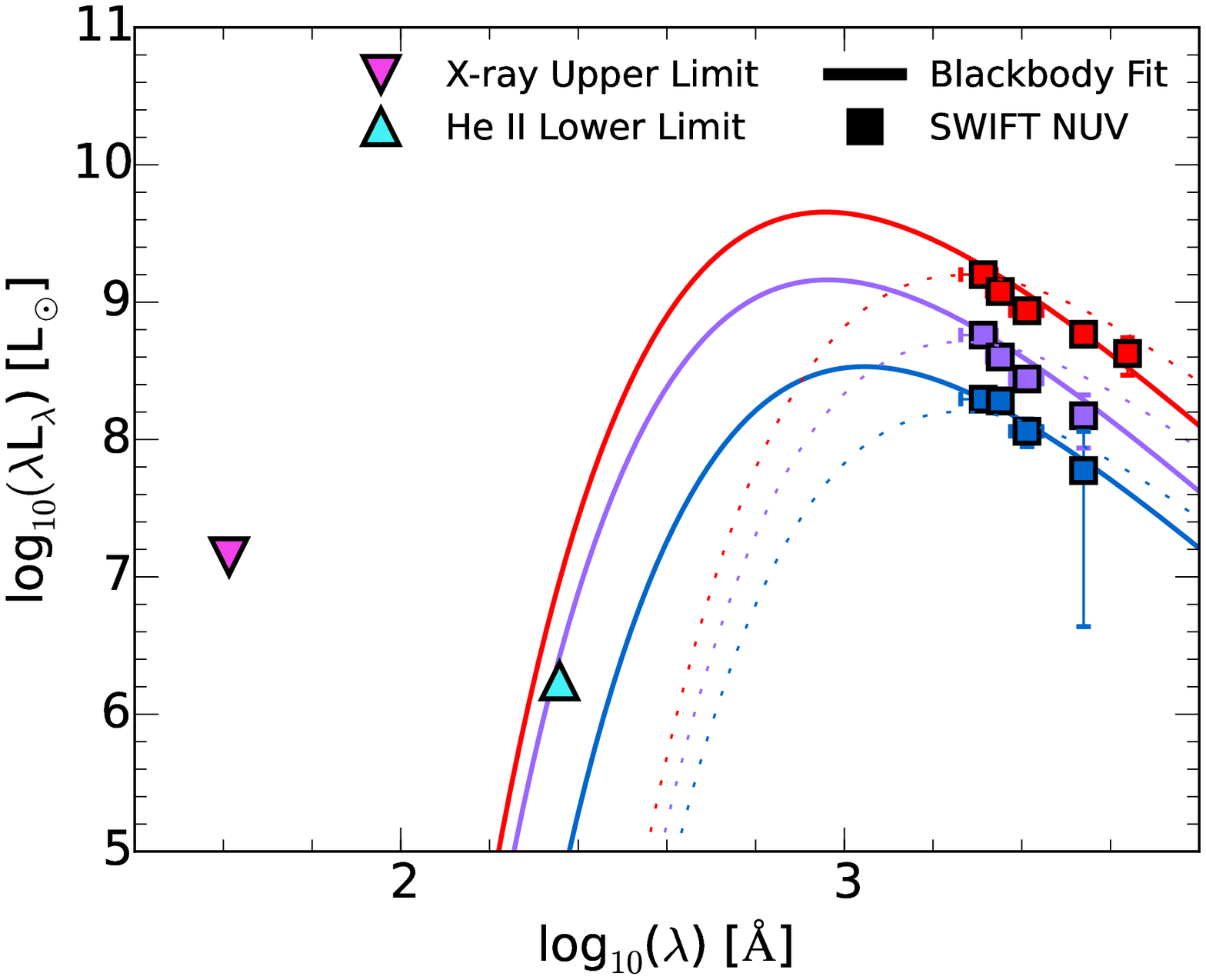}}
\caption{Spectral energy distribution of \fnl. The colored squares show the host-subtracted \SWIFT\ fluxes corrected for Galactic extinction. The solid lines show our blackbody fits to the TDE flux; the dashed lines show similar blackbody models but with the temperature fixed at $T=2\times10^4$~K. The downward and upward arrows show the flux limits inferred from our X-ray and optical spectroscopic observations, respectively.}
\label{fig:sed}
\end{figure}

With a combination of \SWIFT\ and ground-based telescopes, we monitored \fnl\ continuously for the $\sim120$ days following discovery in the near-UV through optical bandpasses. In Figure~\ref{fig:swiftLC} we show the photometric evolution of \fnl\ in the UVOT bands as well as the ground-based observations from Post Observatory. The horizontal dashed lines show the UVOT host magnitudes synthesized from the SED fit to the GALEX (NUV), SDSS ($u'$, $g'$, $r'$, $i'$, and  $z'$), and 2MASS ($J$, $H$, $K_s$) archival data. The vertical marks along the time-axis show the dates corresponding to our spectroscopic observations. The extensive archival data (in particular the GALEX observations) allows us to constrain the host SED quite well. We estimate the accuracy of our synthetic host magnitudes with a bootstrapping scheme in which we perturb the input fluxes by their 1-$\sigma$ errors and fit the resulting SED. We perform 1000 realizations and find that the resulting magnitude estimates are accurate to within $\sim 0.1$ mag, which is much smaller than the UV excess in the bluest UVOT bands at later times. The flux in excess of that the host galaxy decreases as one moves toward redder filters, and the photometry in the optical bands is consistent with the host magnitudes only $\sim20$ days after discovery.

In order to characterize the excess emission, we correct all fluxes for Galactic extinction assuming $R_V = 3.1$ and $A_V = 0.22$ \citep{Odonnell94,Schlafly11}. We then subtract the host flux and model the SED of the flare as a blackbody using MCMC methods \citep{Foreman-Mackey13}. We fit the W2, M2, W1, and U band fluxes, and exclude the B and V photometry due to the negligible contribution of the TDE in those bands. In the later epochs, we only use the W2, M2, and W1 measurements if excess flux in the U band is not significantly detected.

In Figure~\ref{fig:bbTemp} we show the temperature evolution of \fnl\ inferred from the UV/optical photometry (black circles) relative to ASASSN-14ae, ASASSN-14li, and ASASSN-15oi (open blue circles, filled red squares, and open green squares, respectively). There is marginal evidence for temperature modulation during the first $\sim30$ days, but in general the UV/optical continuum is $3-4\times10^4$ K throughout our follow-up campaign. This inference is driven in part by the fact that we lack the wavelength coverage needed to place a strong constraint on the temperature, and so we impose a prior on our fits of $\log(T/K) = 4.5 \pm 0.1$. The temperature evolution is most similar to that of ASASSN-14li, which remained roughly constant, whereas ASASSN-15oi (and ASASSN-14ae to a lesser extent) showed signs of increasing temperature in the weeks following discovery. While the temperature could in principle be higher than our inferred value, the measurements are largely inconsistent with a lower value.

Figure~\ref{fig:bbRad} shows the radial evolution of \fnl\ relative to ASAS-SN TDEs. The dashed line shows the radial evolution of a parabolic orbit with closest approach equal to the tidal radius, where $r_g = G\mbh/c^2$, and we have adopted a fiducial mass $\mbh = 10^7$\msun. Overall, the evolution of the photospheric radius in \fnl\ is not unlike that of ASASSN-14ae and ASASSN-15oi. The characteristic emitting radius of these three objects decreases by a factor of $\sim5-10$ within the first $\sim100$ days, whereas the radius of ASASSN-14li, which remained luminous for much longer than the other ASAS-SN TDEs, decreases only by a factor of $\sim2$. The absolute size of the apparent photospheric radius in \fnl\ is in stark contrast with the more luminous ASAS-SN TDEs. The radius peaks at only a few thousand $\rsun$ and rapidly declines after $\sim10$ days. While we have used the term ``radius'' here, it is unlikely the TDE debris is actually spherically distributed. In any case, it is clear that the emitting surface of \fnl\ is smaller and decreases more rapidly than other optically discovered TDEs.

We estimate the bolometric luminosity by integrating the fits to the SED of the flare for each of our SWIFT epochs, and derive confidence intervals based on the values containing 68\% of the MCMC distribution for each epoch. We show the evolution of the bolometric luminosity as black circles in Figure~\ref{fig:bbLum}. At the earliest phases, \fnl\ is a factor of few fainter than the ASAS-SN TDEs and decreases in luminosity by a factor of $\sim20-30$ over the course of the following weeks. Approximately 50 days after discovery, the decline rate slows, and the emission fades by an additional factor of $\sim2$ between 50 and 100 days after discovery. The combination of a low peak luminosity and rapid decline results in \fnl\ being a factor of $\sim10$ less luminous than the ASAS-SN TDEs after 100 days. 

We do not detect X-ray emission in any epoch, and the proximity of \fnl\ allows us to place relatively strong constraints on the X-ray emission. In order to derive upper limits on the X-ray flux, we adopt a characteristic X-ray temperature of $5\times10^5$~K, as we found for ASASSN-14li \citep{Brown16_14li}. We correct for Galactic absorption assuming a column density of $5.6\times10^{20}$~cm$^{-2}$ \citep{Kalberla05}. The downward arrows in Figure~\ref{fig:bbLum} denote the upper limits on the X-ray emission from \fnl. We perform a similar exercise assuming a $\Gamma=2$ power-law spectrum rather than a blackbody. This provides upper limits which are a factor of $\sim2$ smaller. Given the lack of constraints on the true X-ray spectrum, we adopt the relatively conservative limits derived assuming a blackbody emission spectrum, which are still a factor of $\sim5$ more stringent than those obtained for ASASSN-14ae and ASASSN-15oi. At early times, our limits require that the X-ray luminosity be $<1\%$ of the UV/optical luminosity. This is in stark contrast with ASASSN-14li, in which the X-ray luminosity remained comparable to the UV/optical luminosity for several hundred days \citep{Holoien16_14li,Brown16_14li}.

In Figure~\ref{fig:sed} we show the SED of \fnl\ at the same three epochs shown in Figure~\ref{fig:nuvSpec}, and the color scheme is the same as that used in earlier plots to denote the observation dates. The colored squares show the host-subtracted \SWIFT\ fluxes corrected for Galactic extinction, and the solid lines show our blackbody fits to data. The dashed lines show similar blackbody models but with the temperature fixed at $T=2\times10^4$~K and an increased normalization. In general, the \SWIFT\ fluxes are better described with a relatively hot blackbody rather than the cooler models.

Figure~\ref{fig:sed} also shows constraints on the luminosity at short wavelengths derived from the X-ray limits and the \ion{He}{ii} optical line flux. In order to improve the constraints on the X-ray flux beyond the single epoch limits shown in Figure~\ref{fig:bbLum}, we stack the individual exposures. We obtain a flux limit of $\sim1.1\times10^{-13}$~ergs~s$^{-1}$~cm$^2$ corresponding to a luminosity of $L\sim 1.3\times10^7$L$_{\odot}$, which we show as the downward pointing triangle. Following \citet{Holoien14} we use the flux of the \ion{He}{ii}~$\lambda 4686$\AA\ line to infer a lower limit on the \ion{He}{ii} ionizing luminosity, which we show in Figure~\ref{fig:sed} as the upward pointing triangle. However, this assumes that the material is optically thin to the \ion{He}{ii} emission, and that the \ion{He}{ii} emission arises from a balance of photoionization and recombination. \citet{Roth16} argue that for reasonable physical conditions, the optical depth of the TDE debris in the \halpha\ line can be be significant. Furthermore, given the presence of the collisionally excited resonance lines in the UV, collisional ionization of H may be significant. Thus these assumptions are unlikely to be true for H, but should be more appropriate for \ion{He}{ii}. The \ion{He}{ii} line flux is highly uncertain, but we can still infer an approximate lower limit. Over the course of the observations we measure a \ion{He}{ii} flux corresponding to a luminosity $\gtrsim10^{40}$~ergs~s$^{-1}$ in the early epochs and $\lesssim10^{39}$~ergs~s$^{-1}$ in the later epochs. We adopt a conservative lower limit of 10$^{39}$~ergs~s$^{-1}$ for the \ion{He}{ii}~$\lambda 4686$\AA\ luminosity, and note that decreasing the lower limit by an additional order of magnitude would have no impact on our conclusions. The limits on the X-ray luminosity are quite strong and rule out the presence of a soft ($T\sim5\times10^5$~K) X-ray source down to $\sim10^7$~L$_{\odot}$. On the other hand, the \ion{He}{ii}~$\lambda 4686$\AA\ flux requires a non-negligible amout of flux near 200\AA. Our $\sim3\times10^4$~K blackbody models produce enough ionizing flux to power the \ion{He}{ii} line, while the cooler blackbody models (dashed lines) do not produce enough ionizing flux on their own. However, the X-ray limits are not so stringent as to rule out the presence of a soft X-ray component with enough ionizing flux to power the \ion{He}{ii}~$\lambda 4686$\AA\ line. In fact, the optical line fluxes of other TDEs appear to require an additional source of ionization beyond a simple UV/optical blackbody \citep{Holoien14,Roth16,Holoien16_15oi}. 

There is some tension between the results presented here and those presented in \citet{Blagorodnova17}. First, we systematically favor a higher blackbody temperature ($\gtrsim3\times 10^4$~K instead of $2\times 10^4$~K). This appears to be driven by the assumption in \citet{Blagorodnova17} that the late time flux is dominated by the host galaxy. We find that there is still a significant contribtion to the UV flux from the TDE at late times, as shown in Figure~\ref{fig:swiftLC}. This is corroborated by the fact that the late-time \SWIFT\ W2 flux is $\sim1$ magnitude brighter than the comparable archival \GALEX\ NUV flux. Overestimating the UV flux of the host will naturally drive the temperature estimate downward. We also adopt a slightly higher Galactic extinction ($A_V = 0.22$ instead of $A_V = 0.19$), but this is a secondary effect. The second difference is that \citet{Blagorodnova17} claim an X-ray detection. In our analysis, the \SWIFT\ counts associated with the source are consistent with a background fluctuation, assuming Poisson statistics. Given the low X-ray flux of the detection reported by \citet{Blagorodnova17}, this has no practical impact on our conclusion that the X-rays are not an energetically important component of the observed fluxes.

\section{Conclusions}
\label{sec:conclusions}

We presented for the first time the UV spectroscopic evolution of a TDE using data from HST/STIS. We have shown that the shape and velocity offset of the broad UV emission and absorption features evolve with time. While the UV spectra closely resemble that of ASASSN-14li \citep{Cenko16}, there is little qualitative resemblance to the UV spectra of typical quasars. However, there are some similarities between the of the UV spectra of \fnl, ASASSN-14li, and N-rich quasars \citep{Batra14}, which further suggests that there may be a link between N-rich quasars and TDEs \citep{Kochanek16_stellar}. While the UV line profiles are significantly blended, future higher signal-to-noise observations of brighter TDEs will permit a more quantitative assessment of the line evolution.

The optical spectra of \fnl\ resemble those of several other optically discovered TDEs. Given the emission features in the UV, it seems likely that the anomolous emission features just blueward of \ion{He}{ii}~$\lambda 4686$\AA\ that have been observed in many TDEs are in fact due to \ion{N}{iii} and \ion{C}{iii} Wolf-Rayet blends. These features may contaminate \ion{He}{ii}~$\lambda 4686$\AA, which would make them partly responsible for the large He/H ratios inferred in some TDEs. However, they could also serve as indirect probes of the UV emission when space based observations are not possible. Prompt, high signal-to-noise UV and optical spectra of TDEs are needed before a robust connection between the two can be established.

We monitored the TDE for over 100 days using the \SWIFT\ space telescope and used the XRT and UVOT observations to characterize the broadband SED of the TDE. We showed that the characteristic temperature of the UV/optical emission was $\gtrsim3\times10^4$~K and remained relatively constant over the course of our 100 day observing campaign. Interestingly, the TDE remained dim and faded rapidly relative to other optically discovered TDEs. This is presumably due to an unusually small emitting surface compared to other TDEs. Integrating the UV/optical luminosity over the course of our observing campaign, the total energy radiated by \fnl\ is $\sim5\times10^{49}$~ergs, which is approximately an order of magnitude less than the ASAS-SN TDEs found to date. This corresponds to a small fraction of a solar mass of accreted matter: $\Delta M \sim 3 \times 10^{-4}\eta_{0.1}^{-1}$\msun, where $\eta_{0.1}$ is the radiative efficiency relative to 0.1. Despite the low luminosity of the TDE, we obtain strong limits on the X-ray emission and constrain the 0.3--10~keV X-ray luminosity to be $\lesssim1\%$ that of the UV/optical luminosity. The disparate X-ray properties of \fnl\ and ASASSN-14li are somewhat surprising given the similarities in their UV spectroscopic properties.

Given the low luminosity of \fnl\ it is likely that there are many TDEs of similar (or lesser) luminosity that are simply undetected by current transient surveys. \citet{Kochanek16_demo} argue that if the luminosity function of TDEs scales roughly with the black hole mass function then there should be a large number of relatively low luminosity TDEs around relatively low mass SMBHs that are generally missed by current transient surveys. Given the proximity and low luminosity of \fnl, it appears that this may indeed be the case. As the current generation of transient surveys matures and the detection efficiencies are better characterized, it should be possible to constrain the approximate luminosity function of TDEs and improve the estimate of the volumetric rate. Accurately characterizing the properties of low luminosity events like \fnl\ are a key aspect of determinining the volume corrected statistical properties of TDEs.

\section*{Acknowledgements}

The authors thank Rick Pogge for valuable discussions. 

JSB, KZS, and CSK are supported by NSF grants AST-1515876 and AST-1515927.

TW-SH is supported by the DOE Computational Science Graduate Fellowship, grant number DE-FG02-97ER25308.

BJS is supported by NASA through Hubble Fellowship grant HF-51348.001 awarded by the Space Telescope Science Institute, which is operated by the Association of Universities for Research in Astronomy, Inc., for NASA, under contract NAS 5-26555. 

Support for JLP is provided in part by FONDECYT through the grant 1151445 and by the Ministry of Economy, Development, and Tourism's Millennium Science Initiative through grant IC120009, awarded to The Millennium Institute of Astrophysics, MAS.

S.V.Jr. is supported by the National Science Foundation Graduate Research Fellowship under Grant No. DGE-1343012.

Based on observations made with the NASA/ESA Hubble Space Telescope, obtained at the Space Telescope Science Institute, which is operated by the Association of Universities for Research in Astronomy, Inc., under NASA contract NAS 5-26555. These observations are associated with program \#GO-14781.

This paper used data obtained with the MODS spectrographs built with funding from NSF grant AST-9987045 and the NSF Telescope System Instrumentation Program (TSIP), with additional funds from the Ohio Board of Regents and the Ohio State University Office of Research.

Based on data acquired using the Large Binocular Telescope (LBT). The LBT is an international collaboration among institutions in the United States, Italy, and Germany. LBT Corporation partners are: The University of Arizona on behalf of the Arizona university system; Istituto Nazionale di Astrofisica, Italy; LBT Beteiligungsgesellschaft, Germany, representing the Max-Planck Society, the Astrophysical Institute Potsdam, and Heidelberg University; The Ohio State University, and The Research Corporation, on behalf of The University of Notre Dame, University of Minnesota and University of Virginia.

This research has made use of the XRT Data Analysis Software (XRTDAS) developed under the responsibility of the ASI Science Data Center (ASDC), Italy. At Penn State the NASA \SWIFT\ program is supported through contract NAS5-00136.

This paper uses data products produced by the OIR Telescope Data Center, supported by the Smithsonian Astrophysical Observatory.

Observations made with the NASA Galaxy Evolution Explorer (GALEX) were used in the analyses presented in this manuscript. Some of the data presented in this paper were obtained from the Mikulski Archive for Space Telescopes (MAST). STScI is operated by the Association of Universities for Research in Astronomy, Inc., under NASA contract NAS5-26555. Support for MAST for non-HST data is provided by the NASA Office of Space Science via grant NNX13AC07G and by other grants and contracts.

Funding for SDSS-III has been provided by the Alfred P. Sloan Foundation, the Participating Institutions, the National Science Foundation, and the U.S. Department of Energy Office of Science. The SDSS-III web site is http://www.sdss3.org/.

This publication makes use of data products from the Two Micron All Sky Survey, which is a joint project of the University of Massachusetts and the Infrared Processing and Analysis Center/California Institute of Technology, funded by NASA and the National Science Foundation.

This publication makes use of data products from the Wide-field Infrared Survey Explorer, which is a joint project of the University of California, Los Angeles, and the Jet Propulsion Laboratory/California Institute of Technology, funded by NASA.

This research is based in part on observations obtained at the Southern Astrophysical Research (SOAR) telescope, which is a joint project of the Minist\'{e}rio da Ci\^{e}ncia, Tecnologia, e Inova\c{c}\~{a}o (MCTI) da Rep\'{u}blica Federativa do Brasil, the U.S. National Optical Astronomy Observatory (NOAO), the University of North Carolina at Chapel Hill (UNC), and Michigan State University (MSU).  

\bibliography{ms}
\bsp	
\appendix
\section{Follow-up Photometry}
All UVOT and XRT follow-up photometry is presented in Table~\ref{tab:phot} below. UVOT photometry is presented in the Vega system, and XRT limits are presented in units of $10^{-3}$~counts~s$^{-1}$. The data have not been corrected for Galactic absorption.

\begin{table*}
\begin{minipage}{\textwidth}
\caption{\SWIFT\ Observations.\hfill}\begin{tabular}{cccccccc}
\hline
MJD & XRT Limits [$\times 10^{-3}$ count s$^{-1}$] & W2 & M2 & W1 & U & B & V \\
\hline
57631 & $<$  3.85 & 14.95 $\pm$ 0.04 & 15.25 $\pm$ 0.04 & 15.34 $\pm$ 0.05 & 15.42 $\pm$ 0.04 & 15.97 $\pm$ 0.03 & 15.43 $\pm$ 0.05 \\
57635 & $<$  3.01 & 15.20 $\pm$ 0.04 & 15.42 $\pm$ 0.03 & 15.47 $\pm$ 0.05 & 15.62 $\pm$ 0.05 & 16.02 $\pm$ 0.04 & 15.54 $\pm$ 0.05 \\
57637 & $<$  3.55 & 15.23 $\pm$ 0.04 & 15.48 $\pm$ 0.03 & 15.56 $\pm$ 0.05 & 15.60 $\pm$ 0.05 & 16.12 $\pm$ 0.04 & 15.63 $\pm$ 0.05 \\
57638 & $<$  3.38 & 15.37 $\pm$ 0.04 & 15.59 $\pm$ 0.04 & 15.72 $\pm$ 0.05 & 15.68 $\pm$ 0.04 & 16.05 $\pm$ 0.04 & 15.59 $\pm$ 0.06 \\
57639 & $<$  2.96 & 15.47 $\pm$ 0.04 & 15.69 $\pm$ 0.03 & 15.74 $\pm$ 0.05 & 15.78 $\pm$ 0.05 & 16.13 $\pm$ 0.04 & 15.66 $\pm$ 0.05 \\
57640 & $<$  3.27 & 15.46 $\pm$ 0.04 & 15.72 $\pm$ 0.03 & 15.82 $\pm$ 0.05 & 15.74 $\pm$ 0.05 & 16.14 $\pm$ 0.04 & 15.63 $\pm$ 0.05 \\
57643 & $<$  4.84 & 15.65 $\pm$ 0.05 & 15.87 $\pm$ 0.05 & 15.98 $\pm$ 0.07 & 15.96 $\pm$ 0.07 & 16.23 $\pm$ 0.06 & 15.55 $\pm$ 0.08 \\
57644 & $<$  3.17 & 15.68 $\pm$ 0.04 & 16.02 $\pm$ 0.04 & 16.03 $\pm$ 0.06 & 15.99 $\pm$ 0.05 & 16.22 $\pm$ 0.04 & 15.59 $\pm$ 0.06 \\
57646 & $<$  3.12 & 15.87 $\pm$ 0.04 & 16.14 $\pm$ 0.04 & 16.16 $\pm$ 0.06 & 16.06 $\pm$ 0.05 & 16.29 $\pm$ 0.04 & 15.60 $\pm$ 0.05 \\
57649 & $<$  3.01 & 16.07 $\pm$ 0.04 & 16.28 $\pm$ 0.05 & 16.32 $\pm$ 0.06 & 16.09 $\pm$ 0.05 & 16.28 $\pm$ 0.04 & 15.59 $\pm$ 0.08 \\
57651 & $<$  3.13 & 16.36 $\pm$ 0.04 & 16.55 $\pm$ 0.04 & 16.49 $\pm$ 0.05 & 16.25 $\pm$ 0.05 & 16.30 $\pm$ 0.04 & 15.62 $\pm$ 0.05 \\
57655 & $<$  3.30 & 16.46 $\pm$ 0.06 & 16.71 $\pm$ 0.06 & 16.66 $\pm$ 0.08 & 16.36 $\pm$ 0.08 & 16.35 $\pm$ 0.05 & 15.71 $\pm$ 0.07 \\
57655 & $<$  6.06 & 16.42 $\pm$ 0.07 & 16.71 $\pm$ 0.08 & 16.53 $\pm$ 0.09 & 16.19 $\pm$ 0.09 & 16.34 $\pm$ 0.07 & 15.86 $\pm$ 0.10 \\
57658 & $<$  4.15 & 16.72 $\pm$ 0.05 & 16.85 $\pm$ 0.05 & 16.74 $\pm$ 0.07 & 16.33 $\pm$ 0.07 & 16.33 $\pm$ 0.05 & 15.66 $\pm$ 0.06 \\
57660 & $<$  3.49 & 16.79 $\pm$ 0.05 & 16.96 $\pm$ 0.05 & 16.78 $\pm$ 0.07 & 16.29 $\pm$ 0.05 & 16.39 $\pm$ 0.04 & 15.68 $\pm$ 0.05 \\
57664 & $<$  2.71 & 16.85 $\pm$ 0.05 & 16.98 $\pm$ 0.05 & 16.86 $\pm$ 0.07 & 16.49 $\pm$ 0.06 & 16.30 $\pm$ 0.04 & 15.82 $\pm$ 0.06 \\
57668 & $<$  3.84 & 17.03 $\pm$ 0.06 & 17.13 $\pm$ 0.05 & 16.92 $\pm$ 0.07 & 16.42 $\pm$ 0.06 & 16.40 $\pm$ 0.04 & 15.69 $\pm$ 0.05 \\
57672 & $<$  5.69 & 17.21 $\pm$ 0.08 & \ldots & 17.03 $\pm$ 0.07 & 16.42 $\pm$ 0.06 & 16.31 $\pm$ 0.05 & \ldots \\
57676 & $<$  11.4 & 17.36 $\pm$ 0.07 & 17.28 $\pm$ 0.11 & 17.16 $\pm$ 0.08 & 16.53 $\pm$ 0.07 & 16.35 $\pm$ 0.05 & 15.69 $\pm$ 0.06 \\
57680 & $<$  2.97 & 17.35 $\pm$ 0.06 & 17.38 $\pm$ 0.06 & 17.08 $\pm$ 0.07 & 16.51 $\pm$ 0.06 & 16.41 $\pm$ 0.04 & 15.84 $\pm$ 0.06 \\
57683 & $<$  3.17 & 17.31 $\pm$ 0.07 & 17.38 $\pm$ 0.06 & 17.14 $\pm$ 0.08 & 16.49 $\pm$ 0.07 & 16.38 $\pm$ 0.05 & 15.76 $\pm$ 0.06 \\
57688 & $<$  4.45 & 17.56 $\pm$ 0.07 & 17.55 $\pm$ 0.06 & 17.31 $\pm$ 0.08 & 16.52 $\pm$ 0.06 & 16.41 $\pm$ 0.04 & 15.77 $\pm$ 0.05 \\
57692 & $<$  4.09 & 17.51 $\pm$ 0.07 & 17.51 $\pm$ 0.07 & 17.19 $\pm$ 0.08 & 16.57 $\pm$ 0.07 & 16.43 $\pm$ 0.04 & 15.79 $\pm$ 0.06 \\
57713 & $<$  3.29 & 17.78 $\pm$ 0.06 & 17.78 $\pm$ 0.08 & 17.23 $\pm$ 0.06 & 16.60 $\pm$ 0.05 & \ldots & \ldots \\
57717 & $<$  2.84 & 17.74 $\pm$ 0.06 & 17.81 $\pm$ 0.08 & 17.50 $\pm$ 0.07 & 16.62 $\pm$ 0.05 & \ldots & \ldots \\
57720 & $<$  2.79 & 17.83 $\pm$ 0.06 & 17.70 $\pm$ 0.07 & 17.44 $\pm$ 0.07 & 16.66 $\pm$ 0.05 & \ldots & \ldots \\
57725 & $<$  2.45 & 17.87 $\pm$ 0.06 & 17.72 $\pm$ 0.07 & 17.41 $\pm$ 0.07 & 16.63 $\pm$ 0.05 & \ldots & \ldots \\
57728 & $<$  3.38 & \ldots & \ldots & \ldots & \ldots & \ldots & \ldots \\
57732 & $<$  2.92 & 17.82 $\pm$ 0.06 & 17.72 $\pm$ 0.08 & 17.55 $\pm$ 0.08 & 16.65 $\pm$ 0.05 & \ldots & \ldots \\
57737 & $<$  34.8 & \ldots & 18.13 $\pm$ 0.33 & \ldots & \ldots & \ldots & \ldots \\
\hline
\end{tabular}
\medskip

\raggedright
\noindent Magnitudes and uncertainties are presented in the Vega system. X-ray count rate limits are given in units of 10$^{-3}$ counts per second \\ in the energy range $0.3-10$ keV. Data are not corrected for Galactic absorption.
\label{tab:phot}
\end{minipage}
\end{table*}

\label{lastpage}

\end{document}